\documentstyle[aps,prd,preprint,tighten,floats,epsf]{revtex}

\begin{document}

\preprint{VAND-TH-99-05
\hspace{-30.0mm}\raisebox{-2.4ex}{April 1999}}

\title{The Ubiquitous Inflaton in String-Inspired Models}


\author{Arjun Berera
\thanks{E-mail:
berera@vuhepa.phy.vanderbilt.edu}
and Thomas W. Kephart
\thanks{E-mail: kephartt@ctrvax.vanderbilt.edu}}

\address{
Department of Physics and Astronomy, Vanderbilt
University,
Nashville, TN 37235, USA}

\maketitle

\begin{abstract}

A string theory based inflationary model is developed where the inflaton
interacts with a multitude of massive string level states causing
dissipation of vacuum energy.  Inflation terminates in a warm Universe
without the need for reheating.

\vspace{0.34cm}

\end{abstract}

\medskip

In Press Physical Review Letters 1999

\medskip

hep-ph/9904410

\bigskip


Although inflation has been with us in various forms for two decades now,
it is still a scenario with no real theory.
Strings have been with us for three decades but remain a theory in search
of a particle physics scenario.  Because strings describe the high energy world 
(and likewise high
temperature), they provide the most natural starting point to develop a theory
of inflation.  Up to now string phenomenology has focused on the zero 
mode sector of the theory. However, 
Since the typical string scale is $m_S \sim 10^{17} {\rm GeV}$ 
\cite{Kaplunovsky}, there are many $O(10^2)$ levels \cite{GSW,ftnonehalf} 
below the Planck 
scale $1.22 \times 10^{19} {\rm GeV}$ \cite{ftn3over4}.
A realistic cosmology should not 
{\it a priori}
exclude these higher levels from consideration.

Because of the high density of string states, one generic feature
will be considerable interaction amongst the states and any zero modes
that interact with them.  Thus, if there is an inflaton to be identified
in this regime, associated with its dynamics  will be dissipative effects.
Such a situation provides the basic setting for warm inflation \cite{wi,ab1}.
Warm inflation is comprised of non-isentropic expansion in the
background cosmology \cite{ab2} and thermal seeds of density
perturbations \cite{bf2}.  During warm inflation,
interactions between the inflaton
and other fields cause
the radiation energy
density to remain substantial due to its constant production from
conversion of vacuum energy.  This expansion regime is intrinsically
different from the supercooled inflation regime, since warm
inflation smoothly terminates into a
subsequent radiation dominated regime, without a reheating period.

Recently, a specific quantum field theory realization
of warm inflation has
been developed in terms of
distributed mass models (DM models) \cite{bgr}.  
These models have
been shown to solve the cosmological horizon/flatness
\cite{bgr2} and
scalar density perturbation \cite{abadiab} problems.

Here we will start with the DM model and show that
it naturally generalizes to strings.  {}For this case,
the scales of interest are several orders of magnitude below the
string scale. Nevertheless, the example illustrates
the importance of higher string mass levels.
Although our discussion will be
justified explicitly by this particular model, we will
recognize some robust features of the results that only depend on the
hierarchy of mass levels and not the specific model.

DM models
require many scalar multiplets at 
different mass levels (below but near the Planck scale) interacting 
one after the other with the inflaton $\phi$. Such a hierarchy of mass
levels is naturally suggestive of strings with the inflaton naturally
suggestive of an excited string zero mode that interacts with the massive levels
(or an excited mode at level n interacting with modes at levels higher
than n).
As a first step in realizing this picture, in \cite{bk1} a general SUSY
superpotential was written,
\begin{eqnarray}
W(\Phi,\{X_i\}) & = & 4m {\Phi}^2 + \lambda \Phi^3
\nonumber \\
& + & \sum_{i=1}^{N_M} \left[4\mu_i X_i^2 + f_i X_i^3 +
\lambda'_i \Phi^2 X_i + \lambda^{''}_i \Phi X_i^2 \right].
\label{superpot}
\end{eqnarray}
which represents a chiral superfield $\Phi =(\phi,\psi)$ 
interacting with chiral
superfields $X_i=(\chi_i,\psi_i)$ 
with respective masses $\mu_i$. Within a effective field
theory description, $\Phi$ is the crudest representation of a
string zero mode \cite{ftn1}
interacting with higher mass levels 
$X_i$, in which all symmetries except
SUSY are ignored.  {}For cosmology $\phi$, the bosonic component of $\Phi$,
is interpreted as the inflaton.

When the zero mode is displaced $<\phi> \equiv \varphi \ne 0$,
the interactions in Eq. (\ref{superpot}) alter the masses
of the $X_i$-fields.  {}For thermal field theory this is very important for
it can mean a field $X_i$ that was very heavy $M_i \gg T$ can be made
light due to the effect of $\varphi$.  In particular, DM models
emerge from
shifted coupling arrangements for
$\lambda_i^{\prime}=0$ and $\lambda_i^{\prime \prime} =-2g$
with masses
\begin{equation}
m^2_{\chi_i}= {\rm g}^2(\varphi-M_i)^2 -2gm \varphi -
\frac{3g \lambda}{4} \varphi^2 
\label{mchi}
\end{equation}
and
\begin{equation}
m^2_{\psi_i}= g^2(\varphi-M_i)^2 .
\label{mpsi}
\end{equation}
Irrespective of the mass level scale $M_i$, the $X_i$
field becomes light when $\varphi \sim M_i$ and $\lambda \ll 1$.
In a mass level string interpretation,
if the above shifted coupling arrangement 
is realized for a succession of mass sites $M_i$,
a DM model is realized. In this case, as
$\varphi$ varies over the range of the mass sites, they
successively become light.

In DM models, the effective dynamics of $\varphi$ has two noteworthy
features.  {}First, when the $\chi_i$ and $\psi_i$ -particles are light
$m_{\chi_i,\psi_i} \sim g_i|\varphi-M_i| \stackrel{<}{\sim} T$, the inflaton
can decay into them. This has been shown in 
\cite{hs1,bgr} to lead to an effective viscosity that
acts back on $\varphi$.  {}Furthermore, it has been shown in
\cite{bgr} that this viscosity can be sufficiently large to realize
a slow-roll regime for the inflaton, in which its motion is overdamped.
Second, the distribution of mass sites $M_i$ implies the overdamped
motion can be sustained over quite an extended interval for $\varphi$.

These two features have important consequences for inflationary
cosmology.  Suppose at displaced $\varphi$, a large potential energy
$V(\varphi)$ is being supported.    Overdamped motion then can provide
the concurrent requirements of slow-roll 
$(1/2) {\dot \varphi}^2 \ll V(\varphi)$ and radiation production through
decay of the inflaton into light particles. {}Finally, by having several
mass scales in the shifted mass coupling arrangement that are distributed
over some range, the above process can be sustained sufficiently
long to solve the horizon/flatness problems $N_e > 60$
\cite{bgr2,abadiab}.

The key observation for our example
here is to note that for observationally consistent expansion, $N_e>60$,
the required DM models can be interpreted as arising from a fine
structure splitting of a {\it single} highly degenerate mass level.  
{}For example, for typical cases
studied in \cite{bgr2,abadiab}, it is found that for significant
expansion e-folding, $N_e>60$, if 
$M \approx  g|M_{i+1} - M_i|$ denotes the characteristic
splitting scale between
adjacent levels, warm inflation occurred in the interval \cite{ftn2}
$10^3M \stackrel{<}{\sim} \varphi \stackrel{<}{\sim} 3 \times 10^3M$ 
and of note, at 
temperature \cite{ftn3} 
$M \stackrel{<}{\sim} T$
and not $T$ at the much higher scale of the mass levels
$\sim 10^3M$. What makes these massive
states light is precisely the shifted mass couplings.  In the string
picture, this arrangement corresponds to a fine structure splitting of a
highly degenerate state of very large mass, $\sim M_S$, with 
the fine structure splitting scale several orders of magnitude less than
the mass of the state, say $M \stackrel{<}{\sim} M_{GUT} \sim 10^{-3}M_S$.

Thus the following string scenario suggests itself.  Initially in the
high temperature region, some highly degenerate and very massive level
assumes a shifted mass coupling to $\phi$.  Since all the states in this
level are degenerate, at this point they all couple identically as
$g\sum_i (\phi-M)^2 \chi_i^2$.  The string then undergoes a series of
symmetry breakings that split the
degeneracy and arrange the states
into a DM model $\sum_i (\phi-M_i)^2 \chi_i^2$
with $0 < (M_i-M_{i+1})/M_i \ll 1$.

This string scenario has several appealing features:

\noindent  
(i). {}From our
earlier discussion, strings have an ample supply of highly degenerate
massive states. 

\noindent  
(ii). The generic circumstance is that as
temperature decreases, many of the 
degeneracies will break at least a little, and
for warm inflation a little is all that is needed. Moreover, warm
inflation occurs when $T$ is at or above the fine structure splitting scale
but much below
the scale of the string mass level.  Thus, for the respective mass level,
warm inflation is occurring in a low temperature region. This further
supports the expectation that degeneracies for that level have broken.

\noindent
(iii). The shifted mass coupling to $\phi$ is much more probable to 
occur to a single mass level, albeit highly degenerate, as opposed to the
coincidence probability to several mass levels. 

\noindent
(iv). There are
minimal symmetry requirements for interactions. 
Since zero modes and any higher mass level
modes fall into representations of the gauge and Lorentz groups,
the interactings fields must tensor together to
form singlets.

{}For such a scenario to realize warm inflation, it also requires
the amplitude $\varphi$ remain displaced to sustain a large vacuum
energy density throughout the symmetry breaking scenario, after which
warm inflation must commence. Generically this is a situation in which
this zero mode is trapped in some sort of metastable state. The necessity
of high field amplitude is a probabilistic issue of initial state.
Metastability could arise for
several reasons.  The inflaton could get trapped in a local minima
of a potential as the theory cools. Perhaps more interestingly, as the
theory moves through compactification and the higher mass levels are
transformed to zero modes as the families are, a Kaluza-Klein mode
could get
hung up in a local minima as its initial high field value falls.

Let us now consider this string symmetry breaking 
scenario and estimate the number
of fine structure splittings for the case when a single zero
mode of the string interacts with some mass level n.  Before symmetry
breaking, the massive level is highly degenerate. Assuming SUSY is unbroken, 
part of the
interaction will be represented by the superpotential Eq.({\ref{superpot}),
where the mass parameter $\mu_i=\mu$ is the same for all the states. 
However, this degenerate level is composed of many
irreducible representations (irreps) of the gauge group.  {}For a generic
symmetry breaking scenario,  three types of symmetry breaking can be
considered, compactification (Lorentz symmetry breaking),
SUSY breaking and gauge symmetry breaking.  These 
breakings can occur in
many ways. To focus on a example, suppose compactification has already
occurred and SUSY is unbroken. A standard example is $E_8 \times E_8'$
which after compactification breaks to $E_6 \times E_8'$ with 
three ({\bf 27}) chiral families and the vector like $E'_8$ provides a hidden
sector.  Now gauge symmetry breaking occurs;
suppose $E_6$ breaks \cite{ftn4}
to the standard SU(3) $\times $ SU(2) $\times$ U(1)
model either directly or sequentially through
$\stackrel{v_1}{\rightarrow}$ O(10) 
and $\stackrel{v_2}{\rightarrow}$ SU(5) $\stackrel{v_3}{\rightarrow}$ 
Standard Model. 
The $E_6$ irreps at the higher mass levels will be split into many irreps
of the standard model group. 

Here the VEV's $v_1,v_2,v_3 \ldots$ regulate
the fine structure splitting.
After symmetry breaking at $v_1$, the energy levels will
shift by order this scale,  
depending on details such as 
number of initially degenerate states, symmetries of the states,
size of couplings, and
other parameters in the model. 
{}For a series of symmetry breaking 
$v_1 > v_2 > v_3 \ldots$, the largest symmetry breaking scale,
$v_1$, governs the maximum overall splitting of the level and the
others $v_2,v_3, \ldots$ induce additional hyperfine splittings.

Suppose the n=1 level of the
heterotic superstring is considered. This level has of order
$2\times10^7$ states. Any representation of $E_6$ 
(e.g. {\bf 27},{\bf 78}, $\ldots$) will be split
after symmetry breaking. 
{}For example, an initially
degenerate {\bf 27} splits into 11 finely spaced 
standard model levels (we will ignore all accidental degeneracies).
To go further
it is important to estimate more generally
the degree to which a degeneracy is lifted in
the product of group representations.
A rough estimate of and a lower bound on the lifting of a degeneracy in
breaking a symmetry from a group $G$ to a group  $G^{\prime}$ can be obtained
by considering the group decomposition of irreps
from $G$ to
$G^{\prime}$.
Let $R_1$ and $R_2$ be two irreps of $G$. If $R_1$ and
$R_2$ decompose into $k_1$ and $k_2$ irreps of $G^{\prime}$,
then the product $R_1 \times R_2$ will contain some number $k$ of
irreps which is at least $k_1 \times k_2$. {}For example, let
$G=E_6$ and $G^{\prime} = SU(3) \times SU(2) \times U(1)$ with $R_1={\bf
27}$ and $R_2={\bf 78}$.
It is easy to see that $k_1=11$ and $k_2=25$. Hence
${\bf 27} \times {\bf 78}$
comtains at least $k_1 \times k_2=275$ irreps of
$SU(3) \times SU(2) \times U(1)$.
That this is true can be shown by direct calculation using
${\bf 27} \times {\bf 78}= {\bf 27}+{\bf 351}+ {\bf 1728}$ 
and the fact that ${\bf 351}$ and ${\bf 1728}$ contain
respectively 68 and 320  irreps of $SU(3) \times SU(2) \times U(1)$  for a
total of 399 irreps in their decomposition, i. e., $k_1 \times k_2=275
\le k=399$. Two further cases of note for our later use are ${\bf 27}\times
{\bf 27}=\bar{{\bf 27}}+{\bf 351}+{\bf 351}^{\prime}$
where $k_1 \times k_2=121 \le k=148$ and ${\bf 27} \times
\bar{{\bf 27}} = {\bf 1}+{\bf 78}+{\bf 650}$ 
with $k_1 \times k_2=121 \le k=163$.
A more accurate bound or a direct calculation of $k$ could be gotten
from group projection operators and consideration of $\theta$-series
\cite{conslo},
but the bound we have derived is more than
sufficient for our purposes.

Let us apply our results to the first level of the $E_8 \times
E_8^{\prime}$
heterotic superstring. We assume the 9-dim little group for massive 10-D
states have been reduced on dimensional compactification to 3-dim little
group states via some Calabi-Yau compactification that reduces the gauge
group to $E_6 \times E_8^{\prime}$
This group is then broken by Higgs VEVs 
to $SU(3) \times SU(2) \times U(1) \times E_8^{\prime}$. We find
$k=8360$ Lorentz
singlet irreps of 
$SU(3) \times SU(2) \times U(1) \times E_8^{\prime}$ at this
level with the fine structure splitting scale
$M \sim v_1/k$ where $v_1$ is the largest VEV. {}For a typical GUT
situation
where $v_1 \sim 10^{15}$ GeV, this implies the average mass splitting
between adjacent level 
$M \sim 10^{11}$ GeV.  These numbers are  
in the neighborhood of what is sufficient for the
DM model warm inflation results from \cite{bgr2,abadiab}.
This is an explicit calculation
of the number of irreps.  However, at higher levels we would have to resort to
an estimate from our lower bound. {}From density of states arguments,
we do expect exponential growth in the
number of irreps with level.

The essential feature
exploited in the above example is the prevalence of a high density of
states, which in turn implies interactions and energy exchange are
robust.  These are the elementary ingredients for warm inflation.  One
should anticipate that the above case is indicative of a more general
dynamical structure in strings.  Exploration of this structure may in
turn guide the way to a further refined cosmological dynamics.  To study
this possibility, note two important general features
associated with the higher mass levels of strings.  

\noindent
(i).  String energy levels can change as parameters
(e.g., T,VEVs, etc...) change. When two levels approach each other they can
cross if they correspond to different representations of the symmetry
group. On the other hand, if they are in the same representation, they
will repel and their identities will mix.  This is in exact analogy with
the non-crossing theorem of atomic physics \cite{neuwig}.

\noindent
(ii).  String theory has a Hadedorn temperature \cite{attwit}, 
as typical of any
system with a sufficient rapid growth of its density of states at high
energy \cite{ftn5}.

The first feature suggests that the generic situation 
is likely to be much more rich than the single example given above. 
During any symmetry breaking, mass levels for scalar fields readjust. 
Consider the $n^{th}$ level; as the symmetry breaks, the irreps 
$\{R_n^i\}$ split 
and some states fall, crossing rising states from lower levels 
$R_{n'}^i$ where $n' \le n$. Others rise, crossing states falling 
from above. An orthogonal pair of states $\chi_n$ and $\chi_{n+1}$ 
(from, say, 
level n and n+1) could if uncoupled cross undisturbed, but since in 
general all states will be coupled due to generic terms of the form 
$\lambda {\chi}^2_{n} {\chi}^2_{n+1}$,
crossings will be avoided and mixings will 
occur. 
It has been known 
since the early days of degenerate perturbation theory in quantum 
mechanics that as an energy level moves toward a neighboring level, the 
levels can cross if they correspond to states of different symmetry, but 
if they are states of the same symmetry the crossing is avoided
\cite{neuwig} as the states mix and repel. In this way it is possible to 
transfer energy between states and more generally for dissipation to 
take place.
This scheme seems to be unavoidable, as the string state 
spacing decreases (with $\sqrt{N}$) at higher mass level and the number of 
states grows enormously.  Thus a large number of avoided 
crossings are expected.
All these crossing states will temporarily act as  inflatons 
during their crossing, and many level $n$ states will interact in this 
way with its neighboring levels. 
Scalar VEVs are one of many possible routes to spontaneous symmetry 
breaking. Wilson loops, dynamical symmetry breaking via fermion 
condensates and/or nonperturbative phenomenon all can lead to similar results 
for the generation of effective inflatons and state mixing.

Only one high energy string inflationary scenario was
detailed here, but
it appears to have several appealing features and suggests
some generalizations.  The ultimate question for
assessing this phenomenology is its probability for occurring.
The intrinsic framework of any theory that
explains the cosmological puzzles is probabilistic. This is due to both
intrinsic limitations on attaining observational data and
quantum  mechanical uncertainties that set the initial conditions. 
No theoretical argument can ever conclude that inflation must have
occurred.  Rather, what one can hope to conclude, given a set of
solutions to the cosmological puzzles, is
that inflation is much
more probable than any other explanation.  

Inflationary cosmology is appealing because
there are very few alternative explanations. A
competitive explanations is simply that the initial conditions after the
quantum gravity era just happened to be suitable to evolve into the
observed universe of today, i.e., anthropic principle.  
Compared with this explanation, inflation is
much more probable, since it explains how a very wide class of
initial conditions all evolve into a unique final state
that is suitable to be our observed universe.  

The string inflation interpretation discussed above and the related
quantum field theory analysis in \cite{bgr2,abadiab} are plausible,
and it appears highly probable that a putative inflaton can interact with a
high density of states in the above string inflation picture.
However, identifying the inflaton or inflatons still requires
investigation.

In summary, the high density of string mass level states has been shown
to have important phenomenological consequences for inflationary
cosmology. {}Furthermore, the natural setting is warm inflation, since
the propensity for many states implies considerable interaction and
energy exchange. It is worth noting that the energy scale of warm
inflation scenarios in this picture are generally much below the string
scale, thus compactification most likely has already occurred.  This is
an advantage, since it avoids the problem of inflating unphysical dimensions.
{}Finally, the basic features of this picture are applicable to the recent
low energy string models \cite{lowestng}; 
in fact these models afford more freedom
in adjusting scales.  

This work was supported by the U. S. Department of Energy
under grant DE-FG05-85ER40226.


\begin{references}
\bibitem{ftnonehalf} 
Let us summarize what is known about the spectrum of states
for
the various fundamental string theories\cite{GSW}:
(1) Green-Schwarz $SO(32)$ Open Superstring: The density of states is
$\rho
(m)\sim M^{-9/2}\exp (M/M_{S})$ 
where $M_{S}=\frac{1}{2\pi \sqrt{2\alpha
^{\prime }}}$ and there are 16, 256, 2304, 15360, ... states at the
$0^{th}$
, $1^{st}$, $2^{nd}$, $3^{rd}$, ... levels.
(2) $E_{8}\times E_{8}$ or $SO(32)$ Heterotic Superstring: $\rho (m)\sim
M^{-10}\exp (M/M_{S})$ where $M_{S}=\frac{1}{(2+\sqrt{2})\pi
\sqrt{\alpha
^{\prime }}},$ and there are 7936, 18883584 ... states at the $0^{th}$,
$1^{st}$, ... levels in both these theories.
(3) 26D bosonic: $\rho (m)\sim M^{-25/2}\exp (M/M_{S})$ where
$M_{S}=
\frac{1}{4\pi \sqrt{\alpha^{\prime}}},$ and there are 1, 24, 324, 3200,
25650,
...states at the $-1^{st}$(the tachyon), $0^{th}$, $1^{st}$, $2^{nd}$,
$3^{rd}$, ... levels.
These number densities will be modified when the theory is
compactified to
$4$ dimensions, but the number of states at each level will still grow
sufficiently rapidly that the statements we make below will not be
affected.
\bibitem{ftn3over4} This will be true in almost any
string theory and if we compactify to 4D there are Klusza-Klein 
and winding states as well.
\bibitem{ftn2} In
\cite{bgr2,abadiab} the $\varphi$ interval for warm inflation is
specified in terms of the parameters $i_{min}$ and $N_M$.
\bibitem{ftn3} In \cite{bgr2} $M \sim T$ whereas in
\cite{abadiab} the full range from $M \sim T$ to
$M \ll T$ is treated.
\bibitem{ftn4} The hidden sector $E'_8$ could also break
and radiation would be dissipated in that sector.
However since there are no zero modes (except
$E'_8$ gluons and gluonios) there is a danger of the universe being
matter dominated by $E_8'$ glueballs at $\sim 10^{13} {\rm GeV}$.
\bibitem{ftn5} A string theory is known 
to have a temperature scale $T_H$ where the
partition function for the free string gas fails to converge. It has
been argued  \cite{attwit}
that this behavior is due to
a first order phase transition, a Hagedorn transition. Above $T_H$, the
number of degrees of freedom of the theory has been shown to grow much
more slowly than in any $(d+1)$-dimensional quantum field theory unless
$d=1$. The Hagedorn temperature  coincides with $M_S$ in the bosonic
string and in the type I and II and heterotic superstrings.
In all these cases the free theory has a winding mode whose mass
vanishes at $T_H$ ($T_H$ shifts downward somewhat to the true 
crtical temperature $T_{\rm crit}$, when interactions
are included with the dilaton.).
Above $T_{\rm crit}$ the mode becomes tachyonic, and this is
indicative of a phase transition. The form of the theory changes as a
contribution to the free energy appears at genus zero, where previously
only higher genus contributions existed. While this phase transition is
potentially inflationary, we have limited ourselves to the discussion of
inflation of an effective QFT where $T$ is a few orders of magnitude
below $T_H$. The very interesting regime above $T_H$ corresponds to a
pre-Big Bang (warm) inflation \cite{veneziano}, 
but this is outside the scope of our
considerations here.  We hope to discuss the possibility in a separate
publication. In fact, there are apparently an infinite number of winding
modes whose masses vanish at a sequence of temperatures related to
$T_H$, and this could lead to intriguing results if one is careful on
nearing the regime where statistical definitions begin to fail.
\end{references}
\end{document}